\documentclass[10pt,journal,compsoc]{IEEEtran}
\usepackage{graphicx}
\usepackage{tabularx,booktabs}
\usepackage{multirow}
\usepackage{amsmath}
\usepackage{cite}

\begin{document}

\title{Structuring an unordered text document}

\author{Shashank Yadav,
        Tejas Shimpi,
        C. Ravindranath Chowdary,
        Prashant Sharma,
        Deepansh Agrawal,
        Shivang Agarwal, 
\IEEEcompsocitemizethanks{\IEEEcompsocthanksitem Department
of Computer Science and Engineering, IIT (BHU), Varanasi-221005.\protect\\
E-mail: shashank.yadav.cse16@itbhu.ac.in, shimpitejas.nitin.cse16@itbhu.ac.in, rchowdary.cse@itbhu.ac.in, prashant.sharma.cse16@itbhu.ac.in, deepansh.agrawal.cse16@itbhu.ac.in. shivanga.rs.cse16@itbhu.ac.in}}


\IEEEtitleabstractindextext{%
\begin{abstract}
Segmenting an unordered text document into different sections is a very useful task in many text processing applications like multiple document summarization, question answering, etc. This paper proposes structuring of an unordered text document based on the keywords in the document. 
We test our approach on Wikipedia documents using both statistical and predictive methods such as the TextRank algorithm and Google's USE (Universal Sentence Encoder). From our experimental results, we show that the proposed model can effectively structure an unordered document into sections.
\end{abstract}

\begin{IEEEkeywords}
Ordering sentences, Text processing, Structuring document 
\end{IEEEkeywords}}

\maketitle

\IEEEdisplaynontitleabstractindextext

\IEEEpeerreviewmaketitle

\ifCLASSOPTIONcompsoc
\IEEEraisesectionheading{\section{Introduction}\label{sec:introduction}}
\else
\section{Introduction}
\label{sec:introduction}
\fi
\IEEEPARstart{T}{o} structure an unordered document is an essential task in many applications. It is a post-requisite for applications like multiple document extractive text summarization where we have to present a summary of multiple documents. It is a prerequisite for applications like question answering from multiple documents where we have to present an answer by processing multiple documents. In this paper, we address the task of segmenting an unordered text document into different sections. The input document/summary that may have unordered sentences is processed so that it will have sentences clustered together. Clustering is based on the similarity with the respective keyword as well as with the sentences belonging to the same cluster. Keywords are identified and clusters are formed for each keyword.

We use TextRank algorithm \cite{MihalceaT04} to extract the keywords from a text document. TextRank is a graph-based ranking algorithm which decides the importance of a vertex within a graph by considering the global information recursively computed from the entire graph rather than focusing on local vertex specific information. The model uses knowledge acquired from the entire text to extract the keywords. A cluster is generated for every keyword. 

While generating clusters, the similarity between a sentence and a topic/keyword is calculated using the cosine similarity of embeddings generated using Google's USE (Universal Sentence Encoder) \cite{abs-1803-11175}. USE is claimed to have better performance of transfer learning when used with sentence-level embeddings as compared to word-level embeddings alone. This model is claimed to have better performance even if there is less task-specific training data available. We observed that the quality of clusters/section is better if the similarity of a sentence with the keyword is considered along with the similarity of the sentence with the sentences already available in the respective section.

To test our approach, we jumble the ordering of sentences in a document, process the unordered document and compare the similarity of the output document with the original document.

\section{Related Work}
Several models have been performed in the past to retrieve sentences of a document belonging to a particular topic \cite{Ku:2005:MTD:1076034.1076161}. Given a topic, retrieving sentences that may belong to that topic should be considered as a different task than what we aim in this paper. A graph based approach for extracting information relevant to a query is presented in \cite{ChowdaryK09}, where subgraphs are built using the relatedness of the sentences to the query. An incremental integrated graph to represent the sentences in a collection of documents is presented in \cite{SravanthiCK08,ChowdarySK10}. Sentences from the documents are merged into a master sequence to improve coherence and flow. The same ordering is used for sequencing the sentences in the extracted summary. Ordering of sentences in a document is discussed in \cite{ChowdaryK07}.

In this paper, we aim to generate the sections/clusters from an unordered document. To the best of our knowledge, this is a first attempt to address this problem formally.

\section{Proposed model}
\begin{figure}[!h]
\begin{center}
\includegraphics[scale=.3]{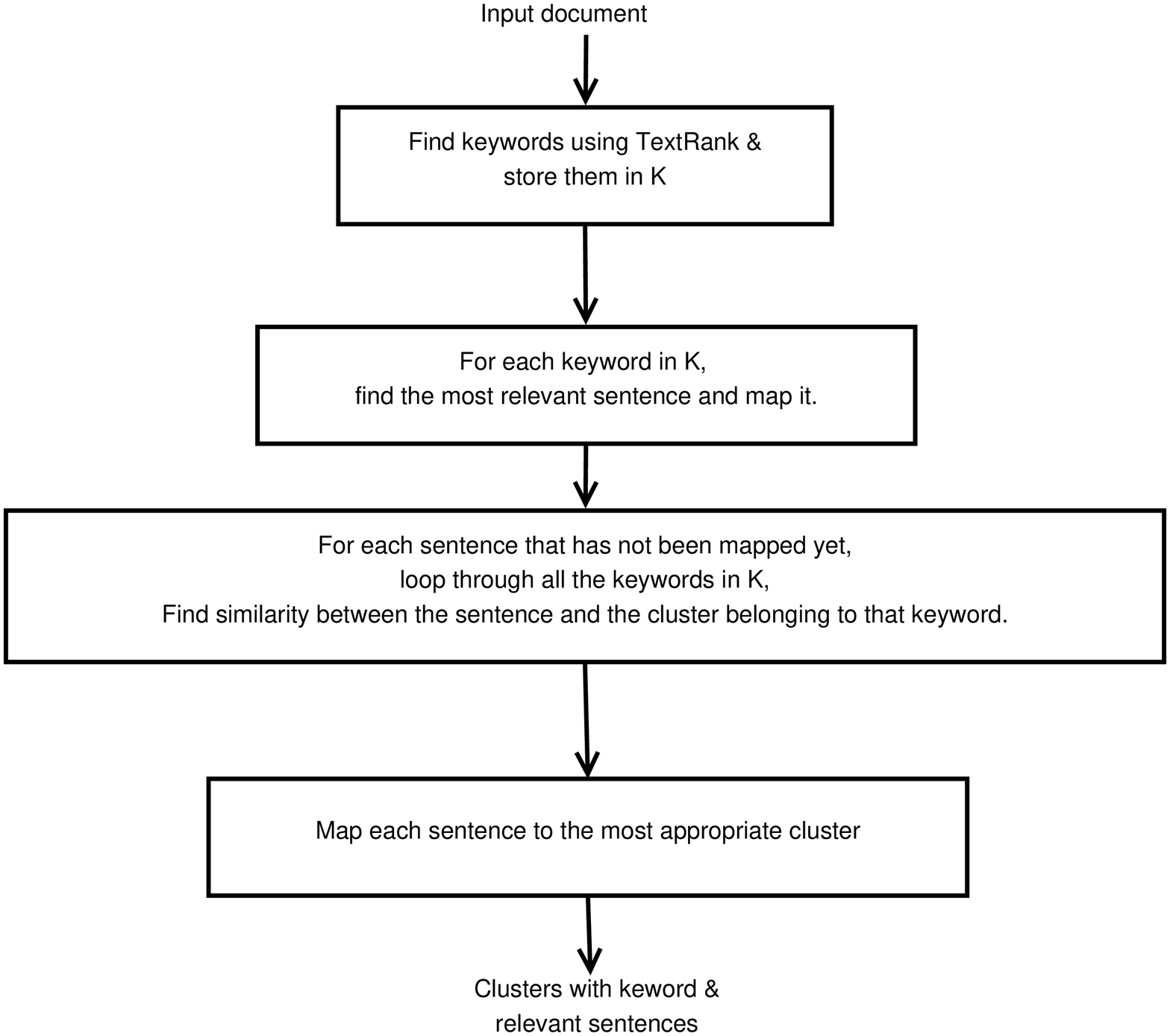}
\caption{Proposed methodology}
\label{fig:figure1}
\end{center}
\end{figure}

Our methodology is described in the Figure \ref{fig:figure1}. The process starts by taking an unordered document as an input. The next step is to extract the keywords from the input document using TextRank algorithm \cite{MihalceaT04} and store them in a list $K$. The keywords stored in $K$ act as centroids for the clusters. Note that, the quality of keywords extracted will have a bearing on the final results. In this paper, we present a model that can be used for structuring an unstructured document. In the process, we use a popular keyword extraction algorithm. Our model is not bound to TextRank, and if a better keyword extraction algorithm is available, it can replace TextRank. 

The next step is to find the most relevant sentence for each keyword in $K$. The most relevant sentence is mapped to the keyword and assigned in the respective cluster. This similarity between the keyword and the sentence is calculated by the cosine similarity of embeddings generated from Google's USE. Now, we have a list of keywords $K$, and a sentence mapped to each keyword. The next step is to map the remaining sentences. In the next step, we go through all the sentences in the document that have not been mapped yet and find the relevant keywords that can be mapped with them. We do this by the following procedure: 

$S_1(x,y)$ is the similarity between the current sentence $x$ and the keyword $y$. $S_2(x,y)$ is the maximum similarity between the current sentence $x$ and the sentences that are already mapped with the keyword $y$. If $y$ has three sentences mapped to it, then the similarity between $x$ and the sentences mapped to $y$ are computed, and the maximum similarity among the three is assigned to $S_2(x,y)$.  
The overall similarity $S(x,y)$ is calculated as:
\begin{equation}
S(x,y)= t*S_1(x,y) + (1-t)*S_2(x,y)
\label{equ}
\end{equation}
We map every other sentence $x$ to the keyword with which they have maximum similarity $Max_y (S(x,y))$. Later, we cluster the keywords along with the sentences mapped to them. The value of $t$ signifies the importance to be given to the keyword and its associated sentences respectively. In our experiments, we empirically fix the value of $t$ to $0.25$.
\subsection{Metrics to evaluate our algorithm}
To evaluate our algorithm, we propose two similarity metrics, $Sim1$ and $Sim2$. These metrics compute the similarity of each section of the original document with all the sections/clusters (keyword and the sentences mapped to it) of the output document and assign the maximum similarity. $Sim1$ between an input section and an output section is calculated as the number of sentences of the input section that are present in the output section divided by the total number of sentences in the input section. To calculate the final similarity (similarity of the entire output document) we take the weighted mean of similarity calculated corresponding to each input section. $Sim2$ between an input section and an output section is computed as the number of sentences of an input section that are present in an output section divided by the sum of sentences in the input and output sections. The final similarity is computed in a similar manner. 
\begin{table}[]
\caption{Results} 
 \centering
 \begin{tabular}{|c|c|c|c|c|}
 
 \hline
    \textbf{Set No.} & \textbf{$Sim1$} & \textbf{$Sim2$} &\textbf{$BaseSim1$} &\textbf{$BaseSim2$}\\
     \hline   
      1 & 0.42625156 & 0.22251266 & 0.35704804 & 0.21033629\\
     \hline
      2 & 0.39003456 & 0.23101839 & 0.33605672 & 0.20739048\\
     \hline
      3 & 0.40847995 & 0.22473745 & 0.34865402 & 0.20686149\\
     \hline
      4 & 0.40035646 & 0.2232119 & 0.34616759 & 0.21629227\\
     \hline
      5 & 0.40785315 & 0.23263196 & 0.32013770 & 0.21621922\\
     \hline
      Average & 0.40664794 & 0.22682247 & 0.34161281 & 0.21141995 \\
    \hline
 \end{tabular}
  \label{tab:Results}
 \end{table}   
\section{Experimental setup and Results}
For our experiments, we prepared five sets of documents. Each set has 100 wiki documents (randomly chosen). Each document is restructured randomly (sentences are rearranged randomly). This restructured document is the input to our model and the output document is compared against the original input document. 

Also, we compare our results with the baseline being the results when we consider only the similarity between sentences and keywords. The results are shown in Table \ref{tab:Results}. Here, both $Sim1$ and $Sim2$ are the mean of similarities of the entire Set. $BaseSim1$ and $BaseSim1$ are computed similar to $Sim1$ and $Sim2$ respectively but with the $t$ value as 1 in Equation \ref{equ}.  It is evident that the results are better if both similarities ($S_1(x,y)$ and $S_2(x,y)$) are considered. 

\section{Conclusions}
We proposed an efficient model to structure an unordered document. We evaluated our model against the baseline and found that our proposed model has a significant improvement. We observed that while ordering an unordered document, the \textit{initial} sentences associated with a keyword/topic play a significant role. 

\bibliographystyle{plain}

\end{document}